\documentclass[prd,twocolumn,showpacs,nofootinbib,superscriptaddress]{revtex4-2}
\usepackage{bm}
\usepackage{indentfirst}
\usepackage{amsmath}
\usepackage{graphicx}
\usepackage{float}
\usepackage{subfigure}
\usepackage{amssymb}
\usepackage{psfrag}
\usepackage{hyperref}
\usepackage{color}
\usepackage{comment}
\usepackage{ulem}
\usepackage{CJK}
\hypersetup{
	colorlinks=true,
	linkcolor=red,
	citecolor=blue,
}

\begin{document}

\begin{CJK*}{UTF8}{gbsn}

\title{Sky localization of space-based gravitational wave detectors}

\author{Chao Zhang (张超)}
\email{chao\_zhang@hust.edu.cn}
\affiliation{School of Physics, Huazhong University of Science and Technology, Wuhan, Hubei 430074, China}
\author{Yungui Gong (龚云贵)}
\email{Corresponding author. yggong@hust.edu.cn}
\affiliation{School of Physics, Huazhong University of Science and Technology, Wuhan, Hubei 430074, China}
\author{Hang Liu (刘航)}
\email{hangliu@sjtu.edu.cn}
\affiliation{School of Aeronautics and Astronautics, Shanghai Jiao Tong University, Shanghai 200240, China}
\author{Bin Wang (王斌)}
\email{wang\_b@sjtu.edu.cn}
\affiliation{Center for Gravitation and Cosmology, Yangzhou University, Yangzhou 225009, China.}
\affiliation{School of Aeronautics and Astronautics, Shanghai Jiao Tong University, Shanghai 200240, China}
\author{Chunyu Zhang (张春雨)}
\email{chunyuzhang@hust.edu.cn}
\affiliation{School of Physics, Huazhong University of Science and Technology,
Wuhan, Hubei 430074, China}

\begin{abstract}
Localizing the sky position of the gravitational wave source is a key scientific goal for  gravitational wave observations.
Employing the Fisher information matrix approximation, we compute the angular resolutions of LISA and TianQin, two planned space-based gravitational wave detectors and examine how detectors' configuration properties, such as the orientation change of the detector plane, heliocentric or geocentric motion and the arm length etc., affect the accuracy of source localization for monochromatic sources at selected values of frequencies. 
We find that the amplitude modulation due to the annual changing orientation of the detector plane helps LISA get better accuracy in the sky localization and better sky coverage at frequencies below several mHz, 
and its effect on TianQin is negligible although the orientation of TianQin's detector plane  is fixed.
At frequencies above roughly 30mHz, TianQin's ability in the sky localization is better than LISA. Further we explore potential space detector networks for fast and accurate localization of the gravitational wave sources.  The LISA-TianQin network has better ability in sky localization for sources with frequencies in the range 1-100 mHz and the network has larger sky coverage for the angular resolution than the individual detector.  
\end{abstract}

\preprint{2009.03476}

\maketitle

\end{CJK*}

\section{Introduction}

Since the first gravitational wave (GW)  event GW150914 observed
by the Laser Interferometer Gravitational-Wave Observatory (LIGO) Scientific Collaboration and the Virgo Collaboration \cite{Abbott:2016blz,TheLIGOScientific:2016agk},
there have been tens of confirmed GW detections by ground-based GW observatories \cite{Abbott:2016blz,TheLIGOScientific:2016agk,Abbott:2016nmj,
	Abbott:2017vtc,Abbott:2017oio,TheLIGOScientific:2017qsa,
	Abbott:2017gyy,LIGOScientific:2018mvr,
	Abbott:2020uma,LIGOScientific:2020stg,Abbott:2020khf,
	Abbott:2020tfl,Abbott:2020niy}.
The ground-based GW observatories can measure GWs within frequency range $10-10^3$ Hz only, and are difficult to explore lower frequency band where a wealth of astrophysical signals reside.  
The proposed space-based observatories including LISA \cite{Danzmann:1997hm,Audley:2017drz},
TianQin \cite{Luo:2015ght} and Taiji \cite{Hu:2017mde}
are expected to detect GWs in the low-frequency regime. 

Accurately localizing GW sources is important, since the source position is correlated with 
 the physical properties of the binary star system
which are necessary to understand the formation and evolution of the binary. 
Moreover, the accurate GW source localization  may provide important information about the environments where such relativistic objects reside. In particular, the accurate knowledge of the GW source position is essential for the follow-up observations of counterparts and the statistical identification of the host galaxy if no counterpart is found, so that we can use GWs as standard sirens to explore the universe's expansion history and understand the problem of Hubble tension \cite{Schutz:1986gp,Abbott:2017xzu,Chen:2017rfc,Wolf:2019hun,Riess:2019cxk}.

One ground-based GW observatory cannot localize GW source, since it is sensible to GW signals from nearly all directions within a few seconds to minutes.  It requires three or more ground-based GW detectors at widely separated sites to locate GW sources with the method of timing triangulation approximation \cite{Fairhurst:2009tc,Fairhurst:2010is,Grover:2013sha}. However, space-based GW detector can measure GWs for months to years, 
the periodic Doppler shift due to the detector motion in space results in amplitude and phase modulations of the detected GW signals which encode information about the detector position and the angular position of the source. 
Therefore, a single space-based GW detector is able to locate the source position.

Like LISA, Taiji is composed of a triangle of three spacecrafts with a larger separation distance in a heliocentric orbit ahead instead of behind the Earth by about $20^\circ$ \cite{Hu:2017mde}. The similarities in configurations construe that similar to LISA, Taiji's angular resolution depends similarly on the type of signal and on how much other information must be extracted.  Similarities and complements between LISA and Taiji imply  that LISA-Taiji network can effectively help to accurately localize GW sources, since  the angular resolution measurements for the network depend on the configuration angle and separation of the two constellations \cite{Ruan:2019tje,Ruan:2020smc}. The network was estimated to improve the angular resolution  over 10 times than each individual LISA or Taiji detector \cite{Wang:2020vkg}.

Unlike LISA and Taiji, TianQin has a geocentric orbit configuration with three spacecrafts orbiting the Earth and further rotating around the Sun together with the Earth. The normal vector of TianQin's detector plane points
to the source RX J0806.3+1527 at $(\theta_{tq}=-4.7^\circ, \phi_{tq}=120.5^\circ)$. TianQin is slightly more sensitive to GWs with higher frequency than LISA.  
The precision of the parameter estimation and the sky localization of
equal mass supermassive black hole binary systems with masses in the range $10^5-10^7M_\odot$
for TianQin was discussed in \cite{Feng:2019wgq}.
It was further argued that the LISA-TianQin network can improve the sky localization of Galactic double white dwarf binaries up to 3 orders of magnitude \cite{Huang:2020rjf}, if compared with single TianQin observation.

For TianQin, considering its special configuration and design, it is important to fully study its angular resolution and uncover its dependencies. It is needed to be clarified in TianQin project how information about the source position encodes in Doppler shift, the translational motion of the detector relative to the source, detector's changing orientation, the rotation period of spacecrafts and the arm length of the detector. 
Different from LISA, whose amplitude modulation improves the sky localization accuracy  below 1 mHz \cite{Cutler:1997ta,Blaut:2011zz},
the detector plane of TianQin points to a fixed direction so there is no amplitude modulation in TianQin which is more sensitive to signals above several mHz. It is interesting to explore how much the sky localization accuracy of TianQin is affected in the absence of the amplitude modulation.
On the other hand, the Doppler modulation becomes stronger as
the frequency of GWs increases and it has the equatorial pattern \cite{Cutler:1997ta,Blaut:2011zz,Zhang:2020drf}, it is natural to ask how the Doppler modulation affects the sky localization
accuracy of TianQin and how to compare it with other space-based detectors at frequencies above 10 mHz. Moreover, there are other questions to be answered, for example:
How the equatorial pattern affects the sky localization accuracy?
Is it possible that TianQin has better sky localization accuracy
than LISA and Taiji even if it has shorter arm length?
Neglecting differences in orbit configurations, if we explore the potential TianQin-LISA network, can it perform better in source localization than Taiji-LISA network?
We will carefully discuss these issues and present sky maps of angular resolutions
for monochromatic sources to identify the sensitive regions of different detectors. These discussions are helpful to improve the detector design.
Throughout the work, we will employ the Fisher information matrix approximation (FIM) to give robust estimation of the sky localization of the source
with a high signal to noise ratio (SNR) where the inverse of FIM
gives the covariance matrix of the parameters.

The  organization of the paper  is as follows.
In Sec. II, we review the FIM method of signal analysis.
In Sec. III, we devise several fiducial detectors to discuss the effects
of different factors on the sky localization. The orbits for these detectors are presented in the Appendix. Then in
Sec. IV we use these results to analyze the angular resolutions of TianQin, compare the result with LISA and finally explore the combined LISA-TianQin network. We present our conclusions and discussions 
in the last section.

\section{Signal Analysis}

For a GW signal
\begin{equation}
h_{ij}(t)=\sum_{A}e^A_{ij}h_A(t),
\end{equation}
the output in the detector $\alpha$ is
\begin{equation}
\label{gwst}
H_\alpha(t)=\sum_A F^A_\alpha h_A(t)+\hat{n}_\alpha(t),
\end{equation}
where $A=+,\times$ stands for the plus and cross polarizations,
$e^A_{ij}$ is the polarization tensor, $\hat{n}_\alpha(t)$ is the detector noise, the angular response function $F^A_{\alpha}$ for the polarization $A$ is
\begin{equation}
\label{faeq1}
F^A_{\alpha}=\sum_{i,j} D^{ij}_\alpha e^A_{ij},
\end{equation}
and $D^{ij}_{\alpha}$ is the detector tensor.
In GR, there are two polarizations. For alternative theory of gravity,
there may exist up to six polarizations \cite{Eardley:1974nw,Liang:2017ahj,Hou:2017bqj,Gong:2017bru,Gong:2017kim,Gong:2018cgj,Gong:2018ybk,Gong:2018vbo,Hou:2018djz,Hou:2018djz}.
In this paper, we consider GR only.
For equal arm space-based interferometric detector with a single round trip light travel,
the detector tensor for a monochromatic GW with the frequency $f$
propagating along the direction $\hat{\omega}$ is
\begin{equation}
D^{ij}_\alpha=\frac{1}{2}[\hat{u}^i_\alpha \hat{u}^j_\alpha T(f,\hat{u}_\alpha\cdot\hat{\omega})-\hat{v}^i_\alpha \hat{v}^j_\alpha  T(f,\hat{v}_\alpha\cdot\hat{\omega})],
\end{equation}
where $\hat{u}_\alpha$ and $\hat{v}_\alpha$ are the unit vectors along the arms of the detector $\alpha$, $T(f,\hat{u}_\alpha\cdot\hat{\omega})$ is \cite{Estabrook:1975,Cornish:2001qi}
\begin{widetext}
\begin{equation}
\label{transferfunction}
%\begin{split}
T(f,\hat{u}_\alpha\cdot\hat{w})=\frac{1}{2}
\{\text{sinc}[\frac{f}{2f^*}(1-\hat{u}_\alpha\cdot\hat{\omega})]\exp[-i\frac{f}{2f^*}(3+\hat{u}_\alpha\cdot\hat{\omega})] +\text{sinc}[\frac{f}{2f^*}(1+\hat{u}_\alpha\cdot\hat{\omega})]\exp[-i\frac{f}{2f^*}(1+\hat{u}_\alpha\cdot\hat{\omega})]\},
%\end{split}
\end{equation}
\end{widetext}
$\text{sinc}(x)=\sin x/x$, $f^*=c/(2\pi L)$ is the transfer frequency of the detector, $c$ is the speed of light
and $L$ is the arm length of the detector.
With the signal $H_\alpha$, we define SNR as
\begin{equation}
\label{snr}
\begin{split}
\rho ^2&=\sum_\alpha \left(H_\alpha|H_\alpha\right) \\
&=4\sum_\alpha\int_{0}^{\infty}df\frac{1}{S_{n,\alpha}(f)} H_\alpha(f)H_\alpha^*(f),
\end{split}
\end{equation}
where the power spectral density $S_{n,\alpha}(f)$ satisfies $\left\langle \tilde{n}_\alpha(f)\tilde{n}_\alpha^*(f)\right\rangle = \frac{1}{2}\delta(f-f^{'} ) S_{n,\alpha}$.
For space-based interferometers, the noise power spectral density $S_n(f)$ is \cite{Luo:2015ght,Hu:2018yqb}
\begin{equation}\label{Sn}
S_n(f)=\frac{S_x}{L^2}+\frac{4S_a}{(2\pi f)^4L^2}(1+\frac{10^{-4}\text{Hz}}{f}).
\end{equation}
For TianQin, the acceleration noise is $\sqrt{S_a}=10^{-15}\
\text{m s}^{-2}/\text{Hz}^{1/2}$,
the displacement noise is $\sqrt{S_x}=1\ \text{pm/Hz}^{1/2}$ and the arm length is $L_t=1.7\times 10^5$ km \cite{Luo:2015ght}.
Its transfer frequency is $f^*_t=0.28$ Hz. 

\begin{figure}
	\centering
	\includegraphics[width=0.8\columnwidth]{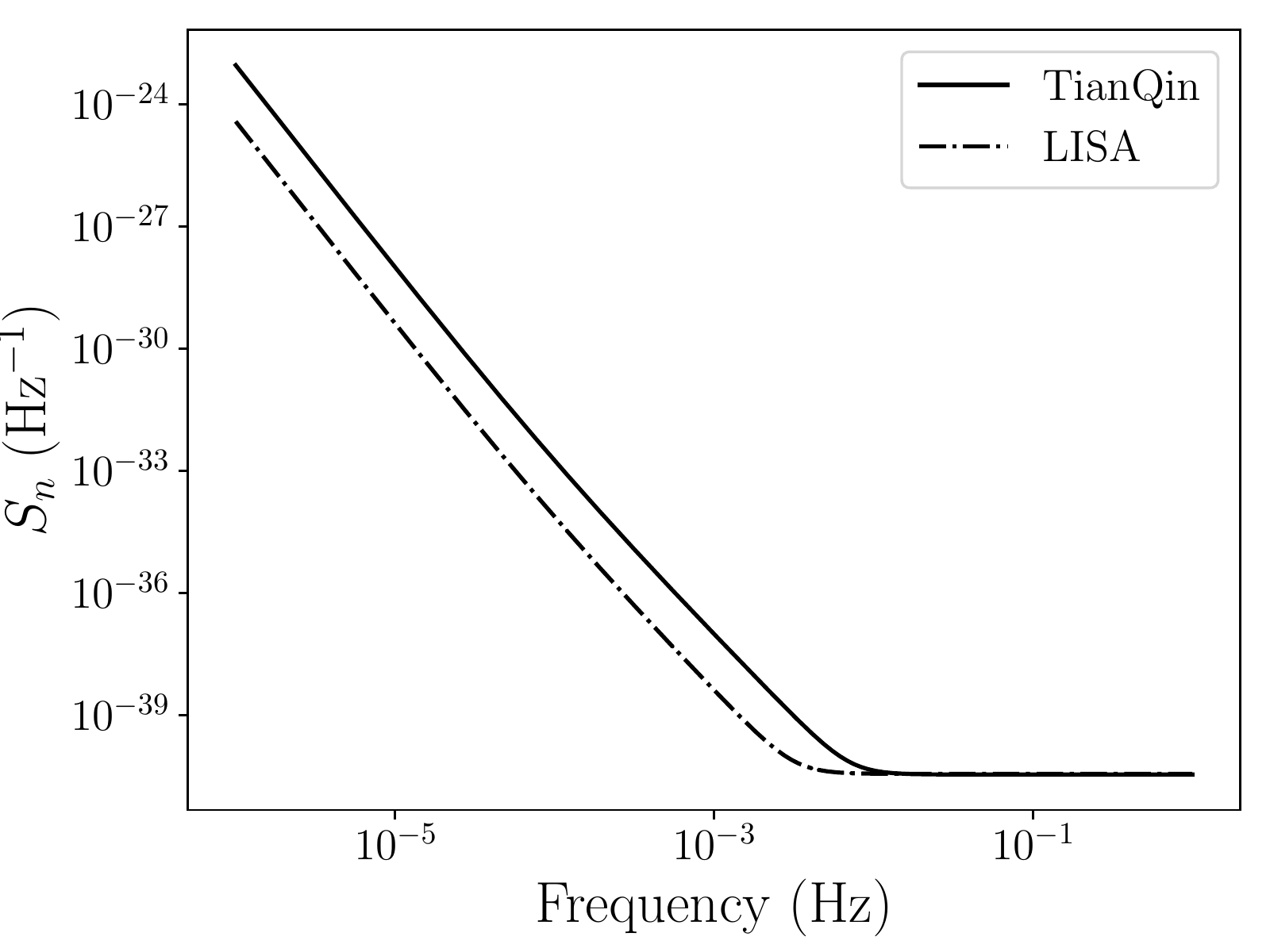}
	\caption{The noise curves for LISA and TianQin.}
\label{noise}
\end{figure}

For LISA, the acceleration noise is $\sqrt{S_a}=3\times 10^{-15}\ \text{m s}^{-2}/\text{Hz}^{1/2}$,
the displacement noise is $\sqrt{S_x}=15\ \text{pm/Hz}^{1/2}$ and the arm length is $L_s=2.5 \times 10^6$ km \cite{Audley:2017drz}.
Its transfer frequency is $f^*_s=0.02$ Hz.
The noise curves for LISA and TianQin are shown in Fig. \ref{noise}.
The ratios of the noise $S_n(f_0)$ between LISA and TianQin are 23 at $f_0=1$ mHz,
1.2 at $f_0=10$ mHz and 0.9 at $f_0=100$ mHz. %(In this paper, we take

Compact binaries containing stellar or intermediate mass black holes, white dwarfs or neutron stars could emit monochromatic GWs
in the early inspiral phase which are detectable by space-based detectors and
the corresponding frequency evolution is negligible during the mission of the detector.
In the lowest order quadrupole approximation,
the two polarizations of monochromatic GWs with the frequency $f_0$ are
\begin{equation}\label{h}
\begin{split}
h_+=&\mathcal{A}\left[1+\left(\vec{L}\cdot\hat{\omega}\right)^2\right]\exp (2\pi i f_0 t + i\phi_0),\\
h_\times=&2i\mathcal{A}\; \vec{L}\cdot\hat{\omega}\;\exp (2\pi i f_0 t + i\phi_0),
\end{split}
\end{equation}
where $\mathcal{A}=2M_1 M_2/(r d_L)$ is the amplitude,
$M_1$ and $M_2$ are masses of the binary components,
$r$ is the distance between them, $d_L$ is the luminosity
distance between the source and the observer, $\vec{L}$
is the unit vector for the binary's orbital angular momentum,
and $\phi_0$ is the initial phase.
We focus mainly on the following seven parameters of the monochromatic GW signal included in Eq. $(\ref{h})$
\begin{equation}
\label{parameters}
    \boldsymbol{\theta}=\{ \theta_s, \phi_s, \theta_L, \phi_L, \mathcal{A}, \phi_0, f_0  \}.
\end{equation}
These parameters describe a monochromatic source by
the source direction ($\theta_s,\phi_s$),
the direction of the binary's orbital angular momentum ($\theta_L,\phi_L$),
the amplitude $\mathcal{A}$, the initial phase $\phi_0$ and frequency $f_0$.
For monochromatic sources there is almost no frequency evolution,
by using the Parseval's theorem \cite{Cutler:1997ta,Vecchio:2004ec},
the FIM becomes
\begin{equation}
\label{tgamma}
\begin{split}
\Gamma_{ij}=&\sum_\alpha \left(\frac{\partial H_\alpha}{\partial \theta_i}\left|\frac{\partial H_\alpha}{\partial \theta_j}\right.\right)\\
=&\sum_\alpha\left[\frac{4}{S_{n,\alpha}(f_0)}\left(\int_{0}^{\infty}\partial_i H_\alpha(f)\partial_j H^*_\alpha(f)df\right)\right]\\
=&\sum_\alpha\left[\frac{2}{S_{n,\alpha}(f_0)}\left(\int_{-\infty}^{\infty}\partial_i H_\alpha(t)\partial_j H^*_\alpha(t)dt\right)\right],
\end{split}
\end{equation}
where $\theta_i$ is the $i$th parameter and $\partial_i H_\alpha=\partial H_\alpha/\partial \theta_i$.
The covariance matrix of the parameters is
\begin{equation}
\sigma_{ij}=\left\langle\Delta\theta^i\Delta\theta^j\right\rangle\approx (\Gamma^{-1})_{ij}.
\end{equation}
For a detected source with a significant SNR  (a threshold of $ \rho\ge 7$),
the angular uncertainty of the sky localization is evaluated as
\begin{equation}
\Delta \Omega_s\equiv2\pi\left|\cos\theta_s\right|
\sqrt{\sigma_{\theta_s\theta_s}\sigma_{\phi_s\phi_s}-\sigma^2_{\theta_s\phi_s}}\,.
\end{equation}

The signal in the detector coordinate system is $H(t)=F^+h_+(t)+F^\times h_\times(t)+\hat{n}_\alpha(t)$. In practice, for space-based detectors,
we often work in the heliocentric coordinate system.
The translational motion of the center of the detector around the Sun leads to an extra phase modulation factor
\begin{equation}
\label{doppler}
    e^{i\phi_D(t)}=e^{2\pi i f_0 R\cos\theta_s\cos\left(2\pi t/T -\phi_s-\phi_{\alpha}\right)/c},
\end{equation}
where $\phi_{\alpha}$ is the ecliptic longitude of the detector $\alpha$ at $t=0$,
the period $T$ of the rotation  is 1 year and the radius R of the orbit is 1 AU.
Therefore, the signal in the heliocentric coordinate system is
\begin{equation}
\label{heliocentricsignal}
H(t)=[F^+(t)h_+(t)+F^\times(t) h_\times(t)]e^{i\phi_D(t)}.
\end{equation}	
Since the phase modulation is very large in most cases
and contains the information about the source location,
this effect is very important for source localization
and it is the same for LISA, Taiji and TianQin.
From Eq. \eqref{doppler}, we get
\begin{equation}
\label{dpfshift}
\begin{split}
      \delta f_0&=\left|\frac{1}{2\pi}\partial_t \phi_D(t)\right|\\
      &=f_0 \cos\theta_s\left|\sin\left(2\pi t/T -\phi_s-\phi_{\alpha}\right)\right| \frac{2\pi R}{c T}\\
      &\sim f_0\frac{v}{c},
\end{split}
\end{equation}
so the periodic phase modulation spreads the measured power of the monochromatic signal of frequency $f_0$ over a range $f_0(1\pm v/c)$ and the spread depends on
the direction of the source.

In addition to the Doppler modulation imposed by the rotation of
the center of the detector around the Sun,
the motion of the spacecrafts with respect to the detector center, the orientation change of the detector plane,
the arm length and the noises of the detector also affect the
accuracy of sky localization.
In the next section,
we analyze how much influences these effects have on
the angular resolution of the detector in detail.

\section{The effect of different constellation on sky localization}

\begin{figure}[htp]
	\centering
	\includegraphics[width=0.8\columnwidth]{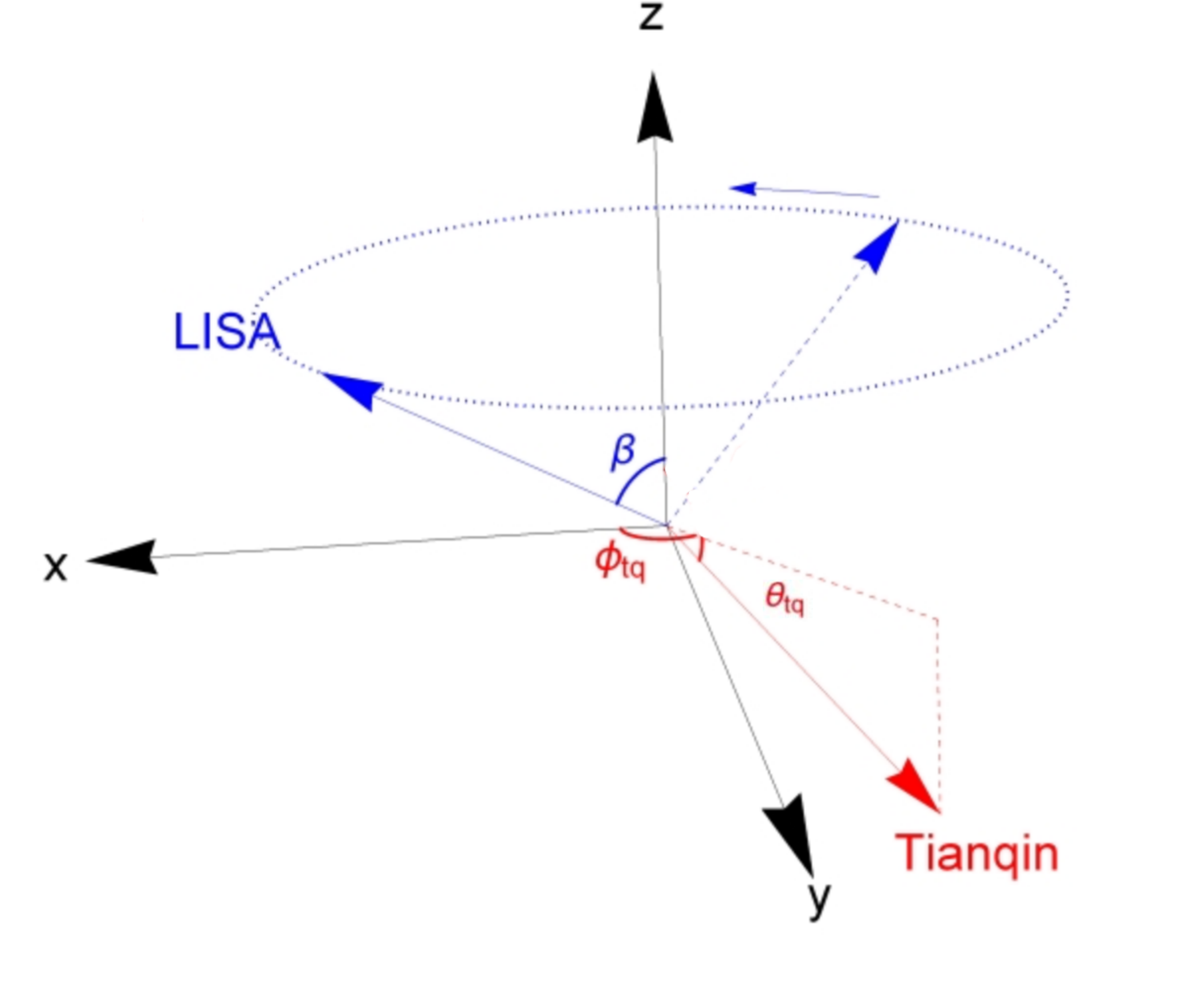}
	\caption{The normal vectors of the detector planes for TianQin and LISA. Illustration from \cite{Liang:2019pry}.}
\label{lstqop}
\end{figure}

\begin{figure}[htp]
  \centering
  \includegraphics[width=0.8\columnwidth]{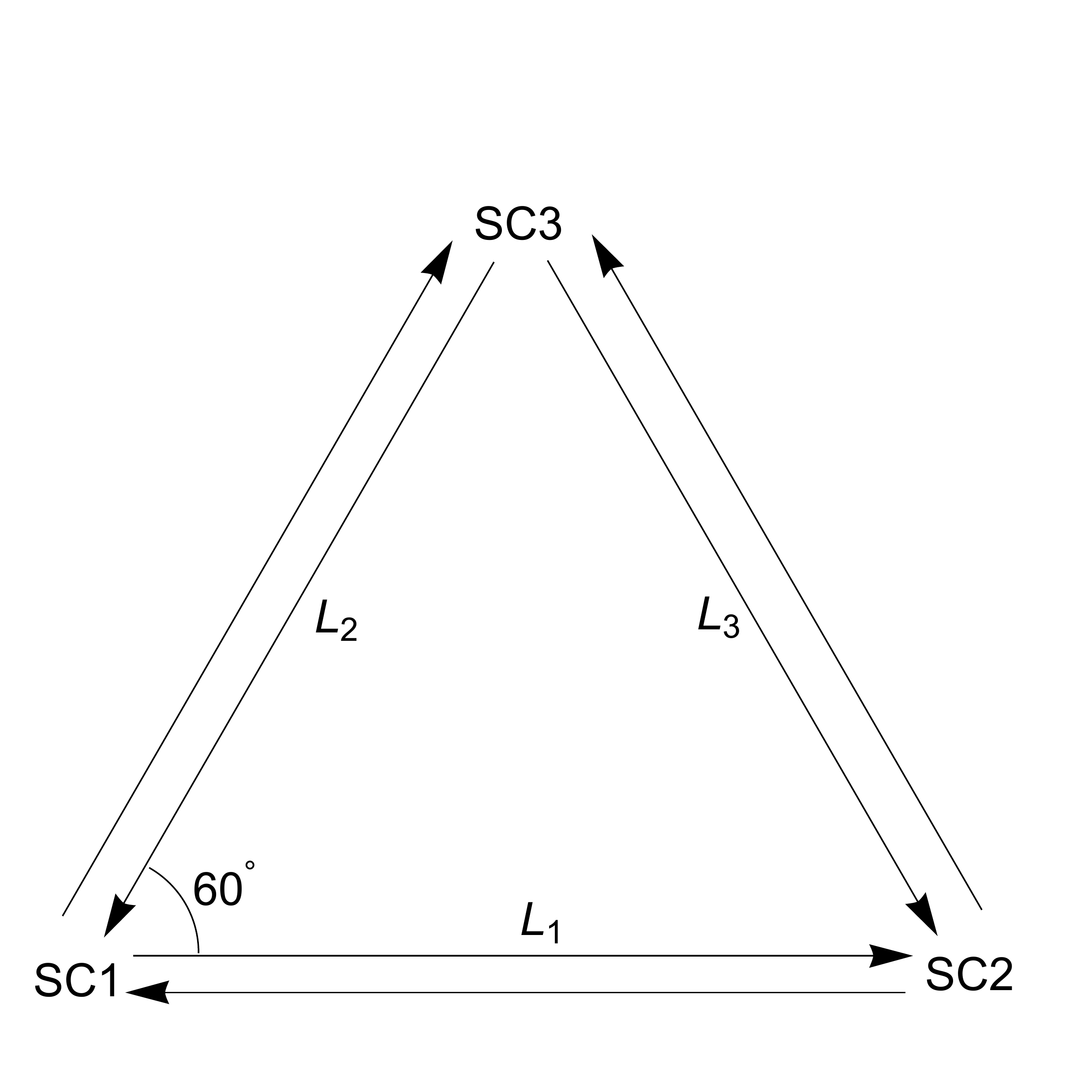}
  \caption{The orientation of three arms in the detector plane.}
  \label{dc}
\end{figure}

LISA mission is proposed as an equilateral triangle constellation with sides of $2.5\times10^6$ km to detect low frequency GWs.
The constellation has an inclination angle of $60^\circ$ with respect to the ecliptic plane and trails the Earth by about $20^\circ$.
The inclination angle ensures the spacecrafts to keep
the geometry of an equilateral triangle throughout the mission.
As shown in Fig. \ref{lstqop}, the normal vector of the detector plane rotates around the normal vector of the ecliptic plane and
forms a cone with $60^\circ$ half opening angle in one year.
Taiji has similar configuration to LISA except that its arm length is $3\times10^6$ km.
TianQin is an equilateral triangle constellation with sides of $1.73\times10^5$ km designed to orbit the Earth with the period of $3.65$ days.
The normal vector of the detector plane points to the direction of RX J0806.3+1527 with the latitude $\theta_{tq}=-4.7^\circ$ and the longitude $\phi_{tq}=120.5^\circ$ as shown in Fig. \ref{lstqop}.
The centers of LISA, Taiji, and TianQin all follow the Earth-like orbit.
Three spacecrafts of the detectors are connected with each other by lasers in a form of equilateral triangle
as shown in Fig. \ref{dc}, we label three spacecrafts as SC1/2/3,
three arms as 1/2/3 and their lengths as $L_1/L_2/L_3$.
We refer to the Michelson interferometer formed  by the arms 1 and 2 as detector I
and the Michelson interferometer formed by the arms 2 and 3 as detector II.
We use both detectors I and II to determine the sky localization, so $\alpha=$ I and II in Eq. \eqref{tgamma}.


As analyzed above, there are four different factors which influence the detectors' localization precision.
To disclose the effect of each factor on the source localization and compare its role in different detectors, we assume the other three factors the same in different detectors.
Because we consider monochromatic sources,
the noise power at the source frequency $f_0$ appears linearly in the Fisher matrix \eqref{tgamma},
so $S_n(f_0)$ can be eliminated by normalizing the result to a fixed SNR \cite{Cutler:1997ta}. In other words, we assume that $S_n(f_0)=1$ for all fiducial detectors
so that the effect of each factor of the constellation on the source localization can be easily understood.
Then we just multiply the ratio of $S_n(f_0)$ to the result when we compare the sky localization of LISA and TianQin.
Before we start to compute the localization accuracy of detectors,
we must give the parameters of sources and ensure that they could be detected.
Thus we fix the SNR of all sources for the detector R (will be defined below) to be 7,
and then we derive the amplitudes $\mathcal{A}$ of sources from this SNR.
This means that the amplitudes of sources being considered are different for different detectors. 
For a complimentary comparison, 
we present the results on the effect of different constellation by taking the same source (i.e., the sources have the same distance and masses) for different constellation in Appendix \ref{results}.
For this setting, perhaps some sources can not be detected by other detectors due to small SNR,
we can increase the amplitudes of all sources by the same factor so that all sources can be detected by all detectors.
Note that the relative angular resolution between detectors remains unchanged although the values of angular resolutions become smaller for each detector.
The other parameters are fixed as $\theta_L=1.0$ and $\phi_L=\phi_0=0$.
The mission time or the observation time $T_o$ is set to be one year.
We simulate 3600 sources uniformly distributed in the sky with $-\pi/2<\theta_s<\pi/2$
and $-\pi<\phi_s<\pi$.
We analyze the source localization accuracy at three frequencies $10^{-3}$ Hz, $10^{-2}$ Hz and $10^{-1}$ Hz.
The three frequencies stand for low, medium and high frequencies relative to the transfer frequency $f^*_s=c/(2\pi L_s)=0.02$ Hz in Eq. \eqref{transferfunction}.

\subsection{The rotation effect}

One major difference between TianQin and LISA/Taiji is the orientation
of the detector plane. Naively we expect that the modulation
of the signal caused by the rotation of the detector plane increases
the accuracy of source localization for LISA and Taiji,
and the precision of sky localization for TianQin becomes less without this modulation effect.
To evaluate the effect of this time-changing orientation of the
detector plane on sources' angular resolutions,
we construct two fiducial GW detectors with the same arm length,
noise curve and rotation period of the spacecrafts.
The first detector is like LISA except that its arm length is $3.7\times10^9$ m, and we call it detector R.
The second detector is like TianQin except that its arm length is $3.7\times10^9$ m and its rotation period around the Earth is one year,
and we call it detector C.
The detailed orbit equations for the detectors R and C are presented in Appendix \ref{orbits}.
For the arm length $L=3.7\times10^9$ m,
the transfer frequency is $f^*=c/(2\pi L)=0.013$ Hz.

\begin{figure}
	\centering
	\includegraphics[width=0.9\columnwidth]{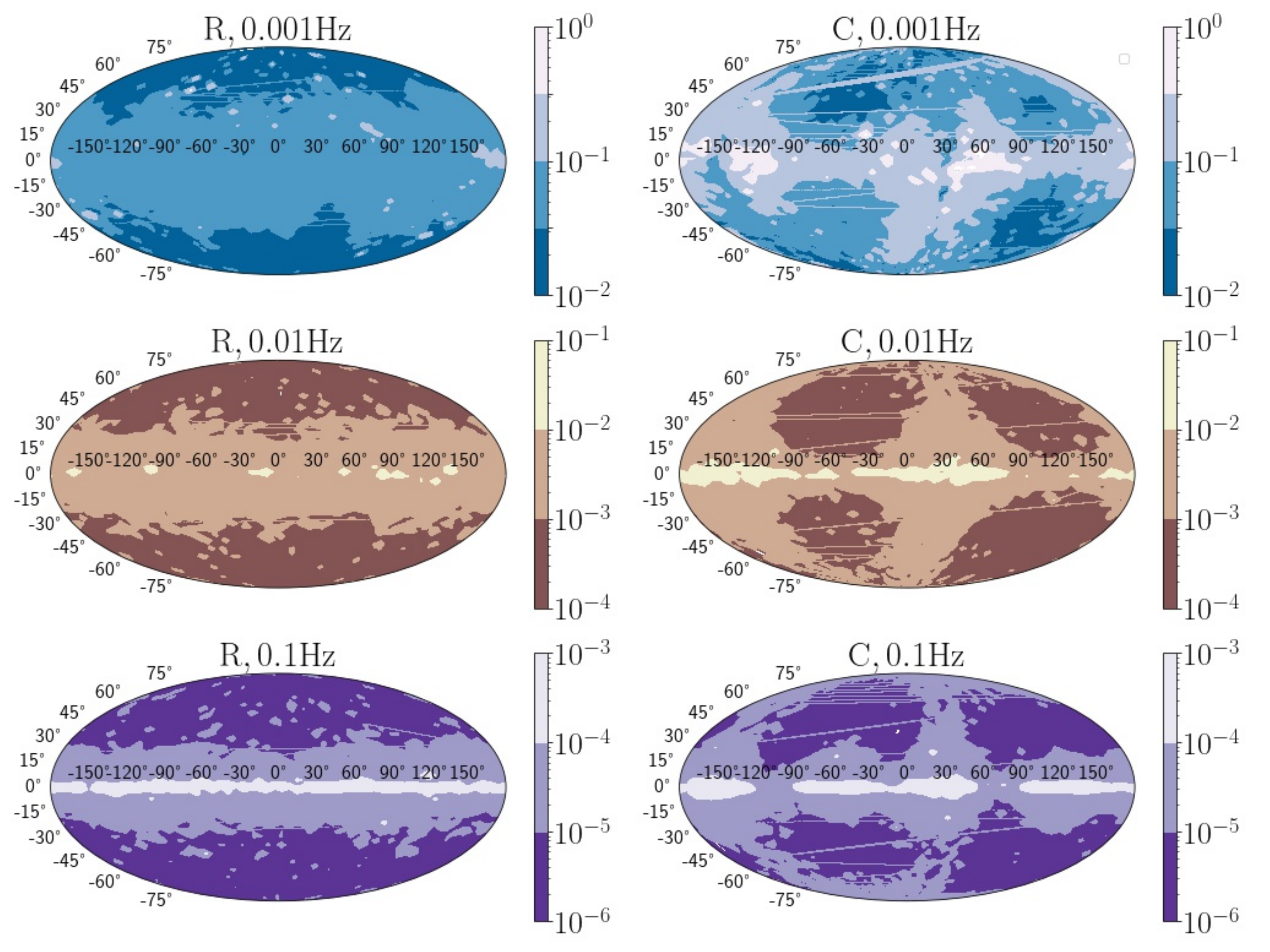}
	\caption{The sky map of angular resolutions $\Delta \Omega_S$ of sources from different directions in the unit
of steradian (1 steradian is 3000 square degrees) for
the fiducial detectors R and C. The horizontal axis represents the longitude $\phi_s$ and the vertical axis represents the latitude $\theta_s$. The left panel is for the detector R and the right panel is for the detector C. From top to bottom, the frequencies of
monochromatic sources are $10^{-3}$ Hz, $10^{-2}$ Hz and $10^{-1}$ Hz.}
\label{rcfig}
\end{figure}

 \begin{figure}
	\centering
	\includegraphics[width=0.9\columnwidth]{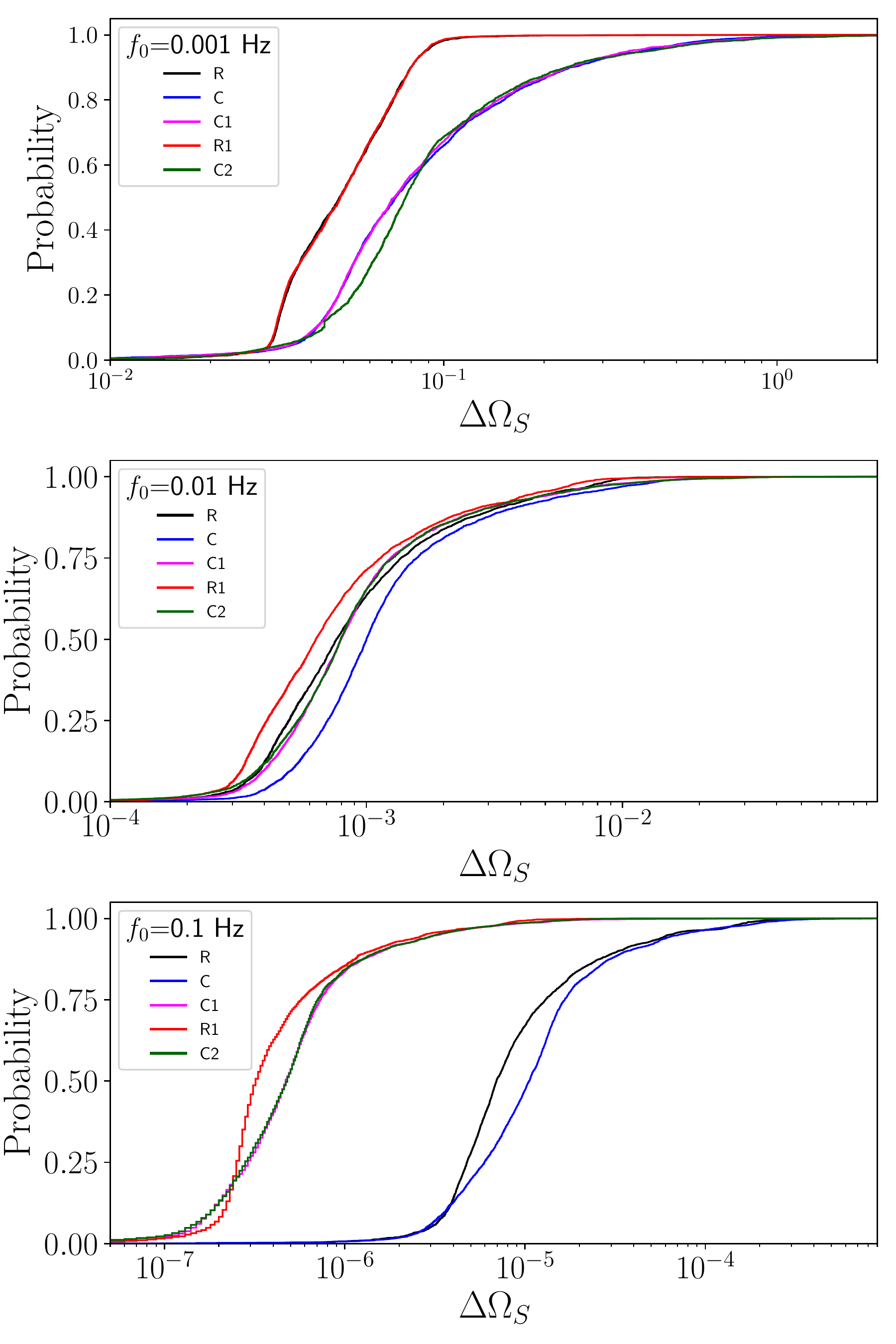}
	\caption{Cumulative histograms of sky localization estimations $\Delta \Omega_S$ for different detectors.}
	\label{hist}
\end{figure}

\begin{figure}
	\centering
	\includegraphics[width=0.8\columnwidth]{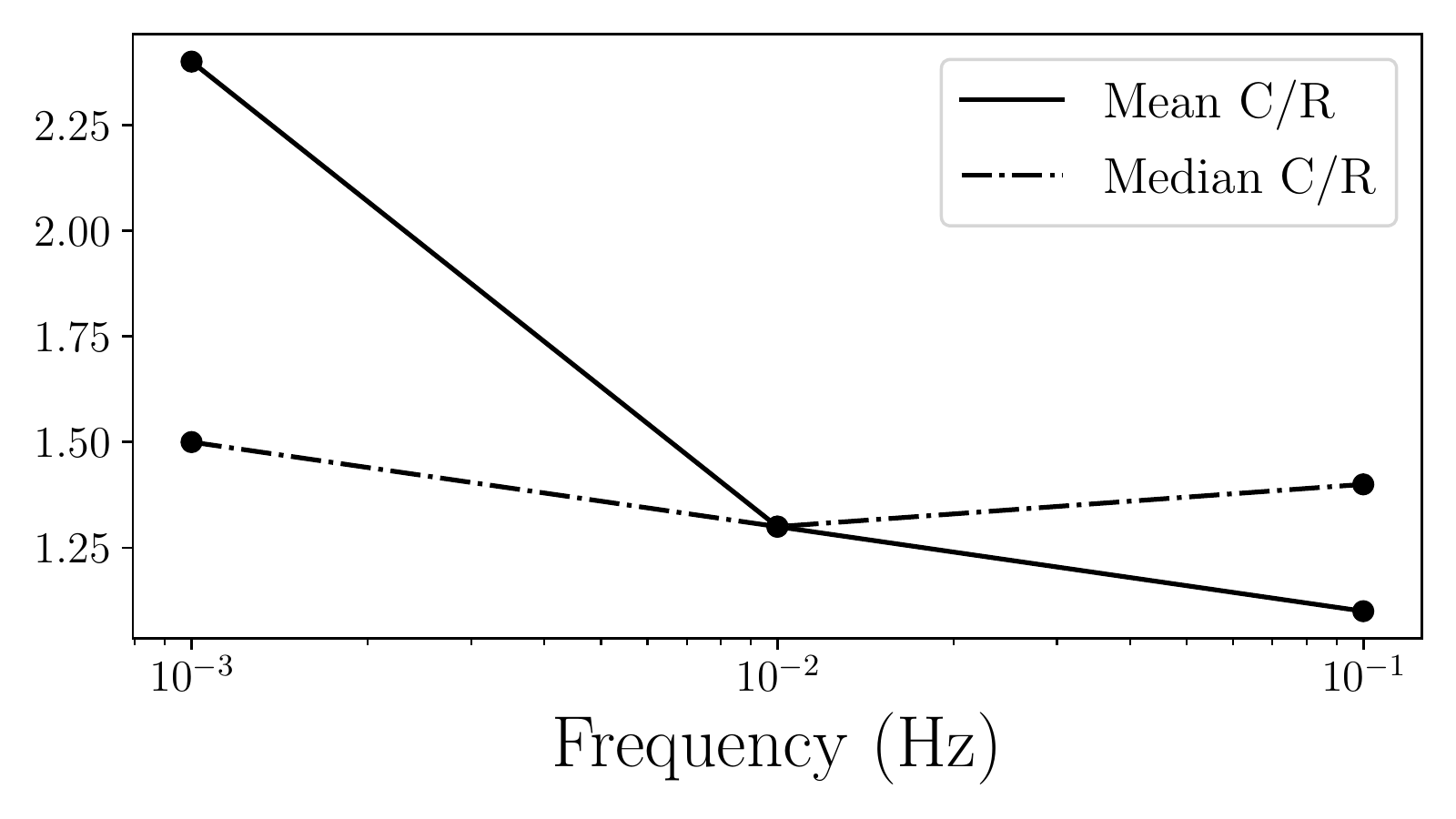}
	\caption{The ratios of the mean and median values of the angular resolutions between the detectors R and C.}
	\label{rcratio}
\end{figure}

The results of angular resolutions at the frequencies $f_0=10^{-1}$ Hz,
$10^{-2}$ Hz and $10^{-3}$ Hz for the detectors R and C are
shown in Figs. \ref{rcfig}, \ref{hist} and \ref{rcratio}.
We show the sky map of the angular resolution for the detectors R and C
in Fig. \ref{rcfig}, the cumulative histograms of the angular resolution in
Fig. \ref{hist} and the ratios of the mean and median angular resolutions
between the detectors C and R
in Fig. \ref{rcratio}.
The mean and median values of angular
resolutions are summarized in Table \ref{meantable}.
From Figs. \ref{hist} and \ref{rcratio},
we see that the effect of the detector's time-varying orientation
decreases as the frequency of GWs increases, the improvement
is only a factor of 2.4 for the mean angular resolution
and a factor of 1.5 for the median angular resolution at the frequency $f_0=1$ mHz with the amplitude modulation caused by the detector's time-varying orientation.
However, the median and mean values of angular resolutions of the detector as usually presented in the literature are incomplete because we don't know where the poorly localized regions are.
Thus we also present the sky map of angular resolutions in Fig. \ref{rcfig}.

 \begin{table}
  \centering
\resizebox{\columnwidth}{!}{
  	\begin{tabular}{|c|c|c|c|c|c|}
		\hline  
$f_0$(Hz)   &R      &C & C1 & C2 & R1 \\ \hline %
&\multicolumn{5}{c|}{Mean Value} \\ \hline
$10^{-1}$   &$2.4\times10^{-5}$     &$2.7\times10^{-5}$      &$1.5\times10^{-6}$      &$1.2\times10^{-6}$     &$1.1\times10^{-6}$  \\ \hline
$10^{-2}$   &$1.8\times10^{-3}$     &$2.3\times10^{-3}$      &$2.0\times10^{-3}$      &$2.1\times10^{-3}$     &$1.6\times10^{-3}$  \\ \hline
$10^{-3}$   &$5.9\times10^{-2}$     &$1.4\times10^{-1}$       &$1.5\times10^{-1}$      &$1.7\times10^{-1}$     &$5.9\times10^{-2}$   \\ \hline
&\multicolumn{5}{c|}{Median Value} \\ \hline %
$10^{-1}$   &$7.0\times10^{-6}$    &$1.0\times10^{-5}$      &$4.7\times10^{-7}$        &$4.8\times10^{-7}$      &$4.2\times10^{-7}$         \\ \hline
$10^{-2}$   &$7.6\times10^{-4}$     &$1.0\times10^{-3}$     &$7.9\times10^{-4}$        &$7.9\times10^{-4}$      &$5.0\times10^{-4}$        \\ \hline
$10^{-3}$   &$4.8\times10^{-2}$     &$7.1\times10^{-2}$     &$7.1\times10^{-2}$        &$7.6\times10^{-2}$      &$4.8\times10^{-2}$          \\ \hline
	\end{tabular}
	}
    \caption{The mean and median values of angular resolutions $\Delta \Omega_S$ for different detectors.}
    \label{meantable}
\end{table}

From Fig. \ref{rcfig}, we see that there are some regions in which the uncertainties of the angular resolutions are large,
but the locations of these regions cannot be seen in the histogram.
As the frequencies of the monochromatic GW sources increase, the
uncertainties of the angular resolutions become smaller.
For the detector R, at the low frequency $f_0=10^{-3}$ Hz,
sky localization uncertainties $\Delta \Omega_S$ are around $0.01$ to $0.1$ steradians across the sky.
Comparing with sources from other directions,
$\Delta \Omega_S$ are more than two
orders of magnitude larger for sources along the equatorial plane with the latitude $\theta_s=0$
in the heliocentric coordinate system at the medium and high frequencies.
For the detector C, the angular resolutions
for sources along the detector plane ($\phi_s$ is around $-150^\circ$ and $ 30^\circ$) and the equatorial plane
in the heliocentric coordinate system are much worse.

In Fourier space, the Doppler effect spreads
the measured power of the monochromatic signal of frequency $f_0$ over a range $f_0(1\pm v/c)$,
where the speed of the motion of the detector's center around the Sun $v\sim 2\pi R/T=3\times10^{4}$ m/s,
so the sky localization is better when $f_0$ increases
and the frequency spread caused by the Doppler effect at
$f_0=10^{-3}$ Hz is $\delta f=f_0 v/c\sim 10^{-7}$.
The Doppler effect on the angular resolution for monochromatic sources is \cite{Blaut:2011zz}
\begin{equation}
\label{omegad}
\begin{split}
\delta\Omega&=\frac{1}{f^2\pi R_s^2 \rho^2\sqrt{1-6x^2}}\left\{\frac{6 R_s^2}{T_o^2}\csc^2(\theta_s)\cos(\theta_s)^2\right. \\
&\left.-12\frac{R_s}{T_0}x \cot (\theta_s) \sin(\phi_s) \csc(\theta_s)+ \csc^2(\theta_s)\right\}^{1/2}\\
&\approx 0.02(1{\rm mHz}/f_0)^2(10/\rho)^2\left|\sin(\theta_s)\right|^{-1},
\end{split}
\end{equation}
where the distance between the Earth and the Sun $R_s=1$ AU, the observational time $T_o=1$ year
and $x=1/\pi$.
It is easy to see that the sky localization becomes better as frequency increases and it depends on the direction of the source.
For sources along the equatorial plane,
the latitude $\theta_s=0$, the sky localization becomes worse, so the Doppler
modulation has the equatorial pattern \cite{Blaut:2011zz}.
On the other hand, the time-varying orientation of the detector's plane spreads the power
over a range $f_0\pm 2/T=f_0\pm6.3\times10^{-8}$ with $T=1$ year 
(i.e., the frequency spread $\delta f=2/T=6.3\times 10^{-8}$),
and it is independent of the source location.
At $f_0=10^{-3}$ Hz, the frequency spread caused by the Doppler modulation is $\delta f\sim 10^{-7}$,
so the frequency spread caused by the amplitude modulation is the same order as that caused by the Doppler modulation,
this explains why the angular resolution for the detector
R is almost the same for sources from any direction,
especially good angular resolutions along the equatorial plane.
As the frequency of GWs increases,
the Doppler effect dominates.
Therefore the angular resolutions for the detector R
and C are almost the same at medium and high frequencies,
and the sky localization becomes worse for sources along the equatorial plane.
For the detector C, if sources are located
along the detector's plane,
then the SNR is smaller and the sky localization becomes worse.

\subsection{The effect of the arm length}
Another difference between TianQin and LISA/Taiji is the distance between spacecrafts.
When the wavelength of GWs is larger than the arm length,
i.e., $f_0<f_*$, the transfer function \eqref{transferfunction}
is almost equal to 1, so the transfer function is independent of
GWs' frequency and the effect of the arm length on the transfer function is negligible. If $f_0>f_*$, then the frequency dependence of the transfer
function deteriorates the response of the detector.
To evaluate the influence of the detector's arm length on angular resolutions,
we devise two more fiducial GW detectors with the same noise curve, rotation period of the spacecrafts and orientation of the detector plane but with different arm lengths.
The third detector is like the detector C except that its arm length is $1.7\times10^8$ m, and we call it detector C1.
For the detector C1, we assume a fiducial Earth with the right mass to provide the orbit. Since the mass
of the Earth does not affect the performance of the detector, so
there is no problem with this assumption for the purpose of the discussion on the effect of the arm length.
The fourth detector is like the detector R except that its arm length is $1.7\times10^8$ m, and we call it detector R1.
For the arm length $L_t=1.7\times10^8$ m, the transfer frequency is $f^*_t=c/2\pi L_t=0.28$ Hz.
The detailed orbit equations for the detectors C1 and R1 are described in Appendix \ref{orbits}. The effect of the arm length can be analyzed by
comparing the angular resolutions for either the detectors R and R1 or C and C1.

\begin{figure*}
	\centering
	\includegraphics[width=0.7\textwidth]{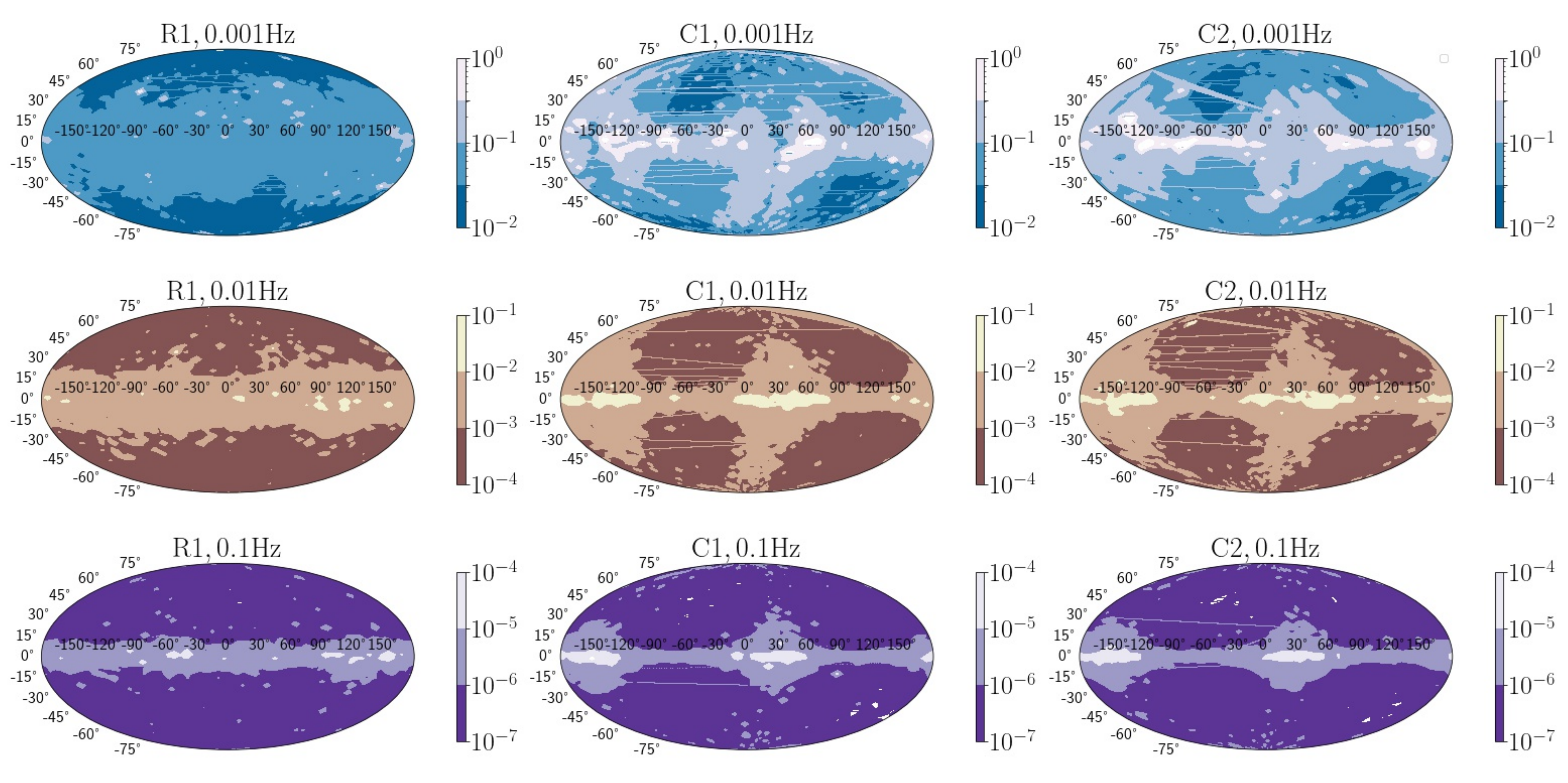}
	\caption{The sky map of angular resolutions $\Delta \Omega_S$ for the detectors R1, C1 and C2. The left panel is for the detector R1,
the middle panel is for the detector C1 and the right panel is for the detector C2. From top to bottom, the frequencies are $10^{-3}$ Hz, $10^{-2}$ Hz and $10^{-1}$ Hz.}
	\label{R1C1fig}
\end{figure*}

\begin{figure}
	\centering
	\includegraphics[width=0.8\columnwidth]{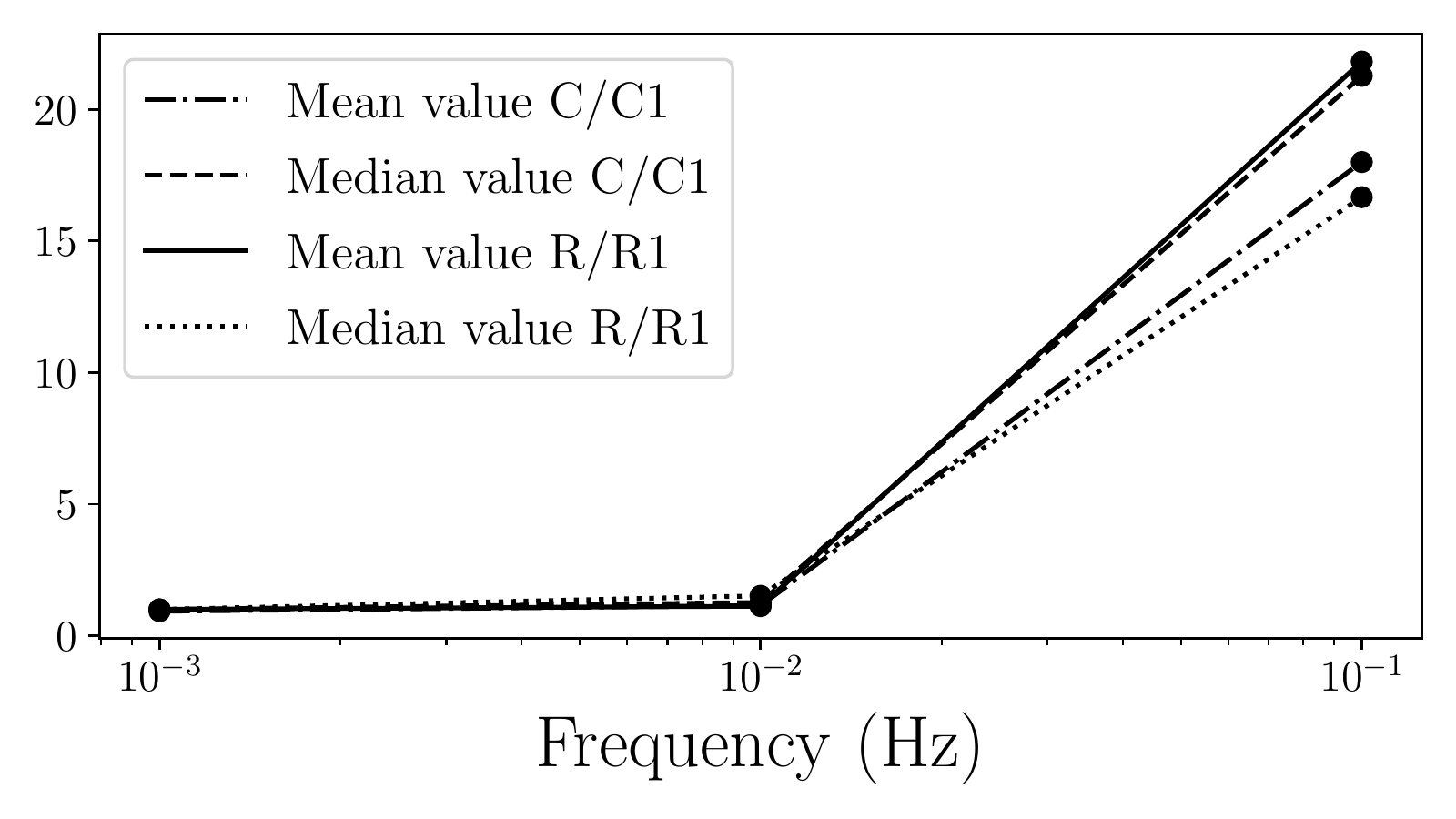}
	\caption{The ratios of the mean and median values of angular resolutions
between the detectors R and R1, and the detectors C and C1.}
	\label{leratio}
\end{figure}

Figure \ref{R1C1fig} shows the sky map of angular resolutions for
the detectors R1 and C1. Because of the Doppler effect, the angular
resolution is better at higher frequency.
Since the frequencies $f_0=1$ mHz and $f_0=10$ mHz
are smaller than the transfer frequencies $f^*=13$ mHz
(for the arm length $L=3.7\times 10^9$ m)
and $f^*_t=0.28$ Hz (for the arm length $L_t=1.7\times 10^8$ m),
so the angular resolutions are similar for either the detectors R and R1
or the detectors C and C1 at $f_0=1$ mHz and $f_0=10$ mHz
as shown in Figs. \ref{hist} and \ref{leratio}.
At the frequency $f_0=0.1$ Hz, the wavelength of GWs is comparable
with the arm length $L_t=1.7\times 10^8$ m and 
less than the arm length $L=3.7\times 10^9$ m,
so angular resolutions of the detectors R1 and C1 are better
than those of the detectors R and C, as shown in Figs. \ref{hist} and \ref{leratio}. We also summarize the mean and median values of angular
resolutions for these detectors in Table \ref{meantable}.
These results tell us that the effect of the arm length is almost the
same for the heliocentric and geocentric constellations
and the factor of improvement on the angular resolution is about $L/L_t=22$ at $f_0=0.1$ Hz.
As expected, TianQin is more sensitive than LISA at frequencies $f_0\gtrsim 10$ mHz.
\subsection{Rotation period of the spacecrafts}
The rotation period of the spacecrafts for TianQin is $3.65$ days
and it is 1 year for LISA/Taiji.
To evaluate the effect of the rotation period of the spacecrafts on sources' angular resolutions,
we devise another fiducial GW detector which is like the detector C1 except that its rotation period is $3.65$ days,
and we call it detector C2.
The detailed orbit equations for the detector C2
are given in Appendix \ref{orbits}. Without specifying the noises,
the detector C2 is the same as TianQin.

Figures \ref{hist} and \ref{R1C1fig} show the histograms and the
sky map of angular resolutions for the detector C2, respectively.
The mean and median values of angular resolutions for the detector C2
are given in Table. \ref{meantable}.
These results show that the influence of the rotation period of the spacecrafts is almost negligible.
Because of the cylindrical symmetry of the rotation, 
the rotations of the spacecrafts  around  the fixed axis have no effect on angular resolutions in one period.

In the real geocentric orbit, the rotation period depends on the arm length.
If we increase the arm length, the detector's sensitivity increases and
the detector is more sensitive at lower frequency.

\section{LISA-TianQin network}

At the frequency $f_0=1$ mHz, the contribution of the 
amplitude modulation due to the time changing orientation
of the detector plane with a period of one year
to the accuracy of the angular resolution
is comparable to that of the Doppler modulation due
to the motion of the center of the detector around the Sun,
so the amplitude modulation helps LISA and Taiji
not only get better angular resolution, but also 
enlarge the sky coverage because the amplitude modulation is
independent of the sources' directions.
At higher frequencies when the wavelength of GWs is larger
than the detector's arm length, the frequency dependent transfer function
deteriorates the SNR registered in the detector,
so we expect that the accuracy of the sky localization for TianQin
is better because of its shorter arm length. 
Now we discuss the sky localization estimations of LISA, TianQin
and the LISA-TianQin network. Since Taiji and LISA have similar constellation,
so we discuss LISA only.
We fix the SNR of all sources for LISA  to be 7,
then we derive the amplitudes $\mathcal{A}$ of sources from this fixed SNR, 
and calculate the angular resolutions of LISA, 
TianQin and the LISA-TianQin network.

\begin{figure}
  \centering
  \includegraphics[width=0.9\columnwidth]{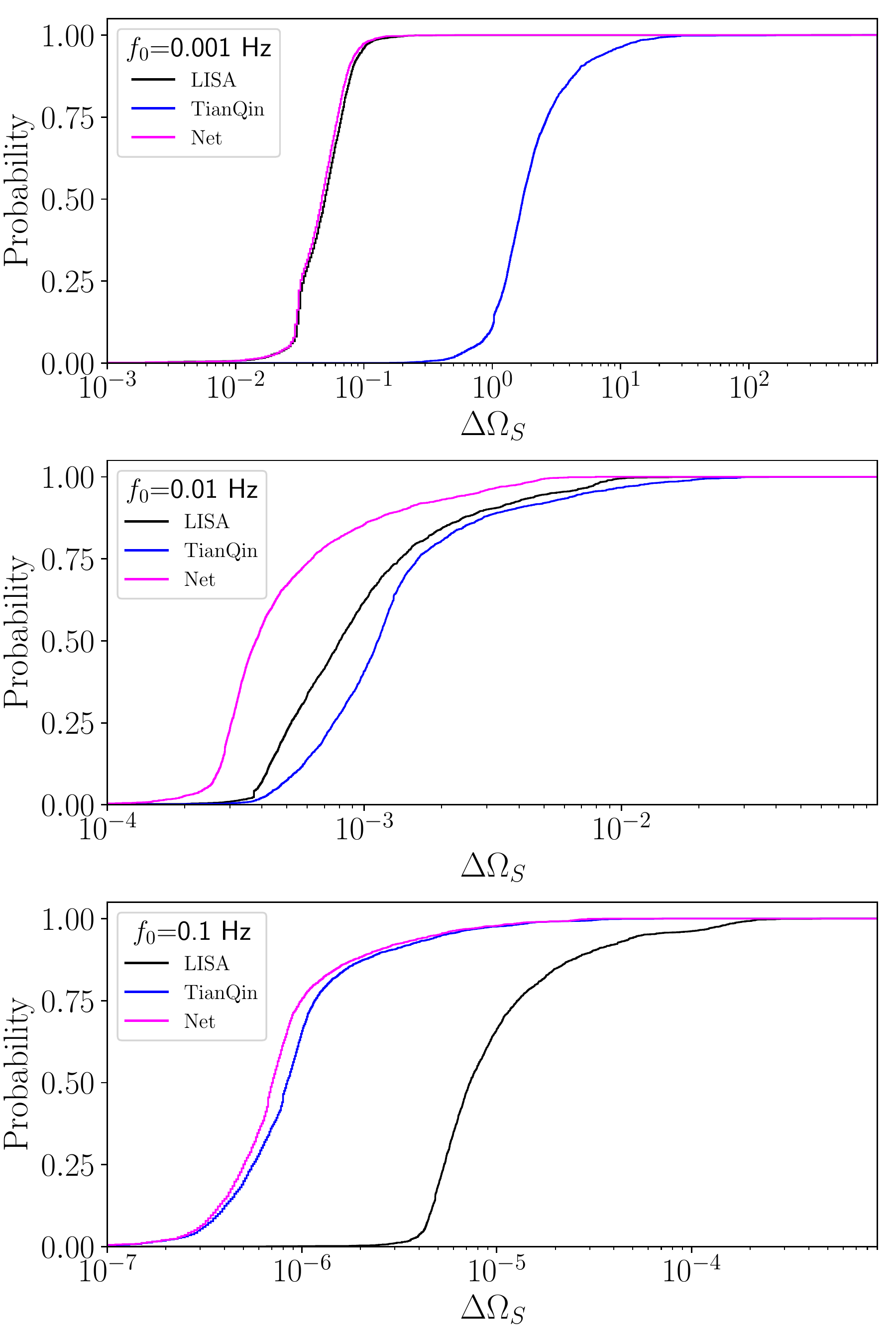}
  \caption{Cumulative histograms of sky localization uncertainties $\Delta \Omega_S$ for LISA, TianQin and their combined network.}
  \label{lstqcomhist}
\end{figure}

We show the histograms of angular resolutions in Fig. \ref{lstqcomhist} and summarize the mean and median values of angular resolutions in Table \ref{lstqcomtable}. 
At the frequency $f_0=10^{-3}$ Hz,
LISA's angular resolution is roughly 50 times better than TianQin
because its noise $S_n(f_0)$
is about 23 times smaller and the amplitude modulation contributes
another factor of 2.4, the angular resolution of the combined network
is almost the same as LISA. 
At the frequency $f_0=10^{-2}$ Hz, LISA's angular resolution is about 1.4 times better than TianQin because TianQin's noise $S_n(f_0)$ is about 1.2 times smaller and its rotation effect is about 1.2 times smaller, 
and the network's angular resolution is a little better than LISA. 
At the frequency $f_0=0.1$ Hz,
the frequency dependent transfer function reduces LISA's sensitivity
by a factor about 10,
so TianQin's angular resolution becomes about $10$ times better than LISA.
Therefore, the angular resolution of the combined LISA-TianQin network 
spans over the frequency ranges 1-100 mHz and it reaches $10^{-6}$
steradians at $f_0=0.1$ Hz. 

\begin{figure*}[htp]
  \centering
  \includegraphics[width=0.7\textwidth]{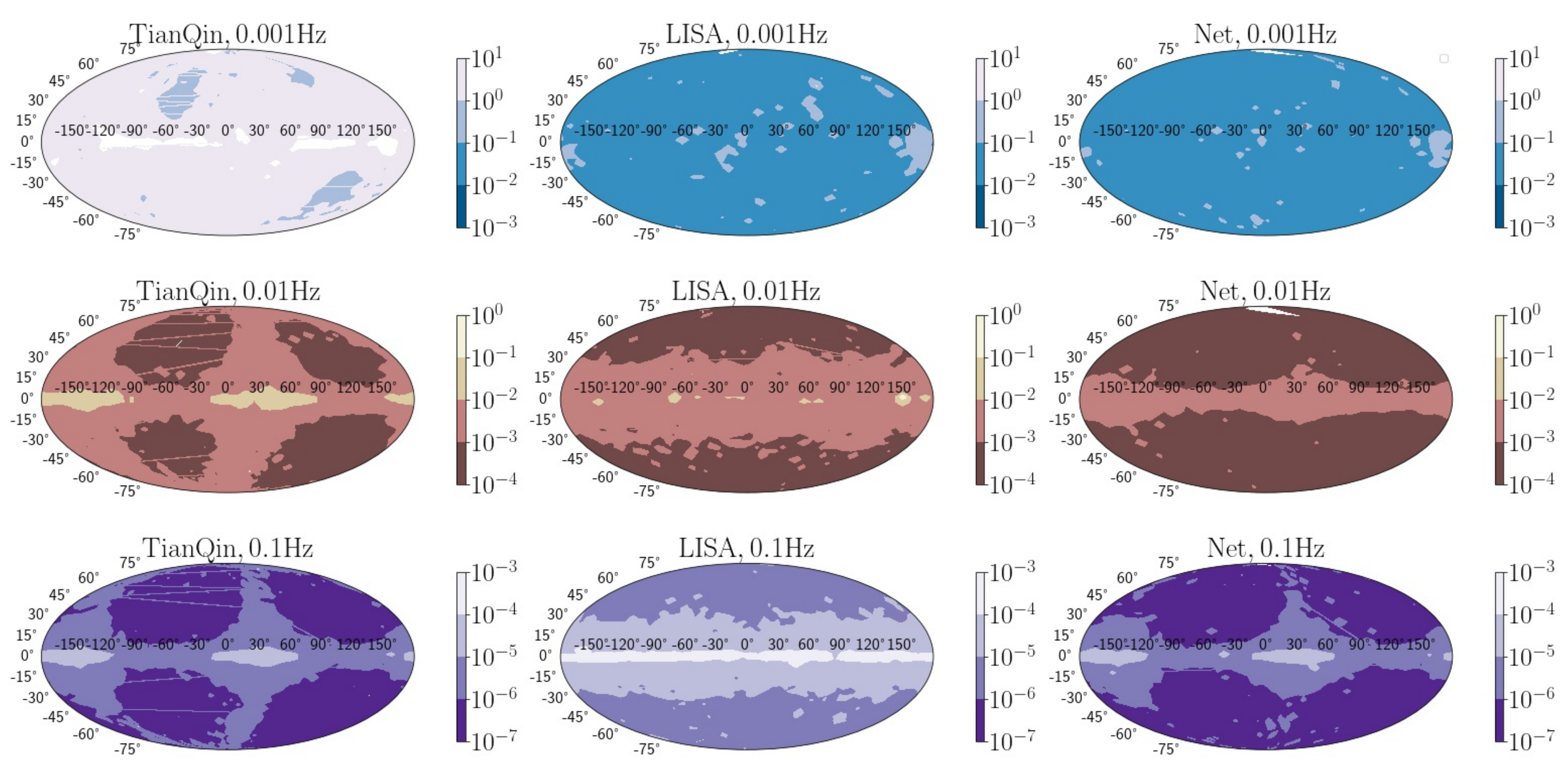}
  \caption{The sky map of angular resolutions $\Delta \Omega_S$ for LISA, TianQin and their combined network at the frequencies $f_0=10^{-3}$ Hz, $f_0=10^{-2}$ Hz and $f_0=10^{-1}$ Hz.}
  \label{tqlisacomall}
\end{figure*}

In Fig. \ref{tqlisacomall}, we plot the sky map of angular resolutions
for LISA, TianQin and their combined network. 
At $f_0=1$ mHz, the amplitude and Doppler modulations contribute
to LISA's angular resolutions, so LISA's angular resolutions
are almost the same across the sky. At higher frequencies,
the contribution of the amplitude modulation is negligible and
only the Doppler modulation matters. For sources along the equatorial plane,
the angular resolution is the worst for both LISA and TianQin
at the frequencies $f_0=10$ mHz and $f_0=100$ mHz.
For TianQin, the worst angular resolution also occurs for sources 
from the directions with $\phi_s=30^\circ$ and $\phi_s=-150^\circ$.
Fig. \ref{tqlisacomall} shows that the combined network enlarges the sky coverage
in addition to the slight improvement on angular resolutions at $f_0=10$ mHz and $f_0=100$ mHz.

\begin{table*}[htp]
\centering
\resizebox{\columnwidth}{!}{
 	\begin{tabular}{|c|c|c||c|c||c|c|}
	\hline
	  &\multicolumn{2}{|c||}{LISA}  & \multicolumn{2}{|c||}{TianQin}                & \multicolumn{2}{|c|}{Network} \\\hline
	$f_0$(Hz) & Mean & Median & Mean & Median & Mean & Median\\\hline
	$10^{-1}$   &$2.3\times10^{-5}$  &$7.4\times10^{-6}$    &$2.1\times10^{-6}$      &$8.3\times10^{-7}$   &$1.8\times10^{-6}$                &$7.1\times10^{-7}$  \\ \hline
    $10^{-2}$   &$2.0\times10^{-3}$  &$8.2\times10^{-4}$     &$2.5\times10^{-3}$      &$1.1\times10^{-3}$   &$8.9\times10^{-4} $              &$3.8\times10^{-4}$  \\ \hline
    $10^{-3}$   &$6.2\times10^{-2}$  &$5.03\times10^{-2}$     &$3.7$                   &$1.7$                &$5.7\times10^{-2}$               &$4.9\times10^{-2}$  \\ \hline
\end{tabular}
}
\caption{The mean and median values of angular resolutions for monochromatic sources.}
\label{lstqcomtable}
\end{table*}

\section{Conclusion}

The detector's noises, arm length, time-changing orientation, motion around the Sun
are four main factors that influence the accuracy of source localization
for spaced-based GW observatories. 
The detector's noises and arm length affect the angular resolution through SNR.
The Doppler modulation on the amplitude and phase of the GW signal 
imposed by the translational motion of
the detector's center around the Sun carries the position information of the source.
For LISA and Taiji, the amplitude of GW signal is further modulated by the annual rotation of the detector's orientation.
Both the Doppler and amplitude modulations 
spread the power of the monochromatic GWs over a frequency range in Fourier space.
The amplitude modulation improves the accuracy of the sky localization
below several mHz and the improvement is the same for sources from
different directions.
The effect of the Doppler modulation on the angular resolution
becomes better as the frequency of the monochromatic GW source increases,
and this effect depends on the direction of the source.
For sources along the plane with $\theta_s=\pi/2$ in
the heliocentric coordinate system, the improvement by the
Doppler modulation on the angular resolution is the largest.
For frequencies above approximately 5 mHz,
the contribution of the amplitude modulation is negligible and
only the Doppler modulation matters. 
Therefore, TianQin's angular resolution is not affected much due to the lack of amplitude modulation.
By pointing to the specific source, TianQin has some blind spots, 
its angular resolution is not good 
for sources from the directions with $\phi_s$ around $30^\circ$ or $-150^\circ$.

At the frequency $f_0=1$ mHz, the contribution of the
amplitude modulation is comparable to that of the Doppler modulation,
so the amplitude modulation helps LISA and Taiji
not only get better angular resolution, but also
enlarge the sky coverage.
At higher frequencies when the wavelength of GWs is larger
than the detector's arm length, the frequency dependent transfer function
deteriorates the SNR registered in the detector.
For the monochromatic sources with the frequency $f^*_s=c/(2\pi L_s)<f_0<f^*_t=c/(2\pi L_t)$,
TianQin's angular resolution is better than LISA's by 
a factor of roughly $S_{n,t} L_s/(S_{n,s} L_t)$, here
$S_{n,s}$ and $L_s$ are LISA's noise curve and arm length,
and $S_{n,t}$ and $L_t$ are TianQin's noise curve and arm length.
Since LISA and Taiji have the best sensitivity at around 1 mHz,
and TianQin is more sensitive at 10 mHz, the LISA-TianQin or TaiJi-TianQin network
have better ability of sky localization for sources in the frequency range 1-100 mHz and the network has larger sky coverage for the angular resolution than the individual detector.
By assuming SNR=7 for LISA, the network's
angular resolution is about 200 square degrees at 1 mHz, 
3 square degrees at 10 mHz and 0.005 square degrees at 100 mHz.
The understanding of the effects of the amplitude and Doppler modulations
on LISA/Taiji/TianQin can be used to optimize their constellations. 
In particular, the result may help TianQin to improve the design so
that the equatorial pattern and the blind spots can be avoided.

\begin{acknowledgments}
This research was supported in part by the National Key Research and Development Program of China under Grant No. 2020YFC2201504,
the National Natural Science
Foundation of China under Grants No. 11875136 and No. 12075202, and
the Major Program of the National Natural Science Foundation of China under Grant No. 11690021.
\end{acknowledgments}

\appendix
\section{DETECTOR'S ORBITS}
\label{orbits}
\subsection{The orbits for the detectors C, C1, and C2}
In the ecliptic coordinate system, the orbit  $\vec{r}^T_n(t)=(X^{T}_n(t),Y^{T}_n(t),Z^{T}_n(t))$, ($n=1,2,3$)
is
\begin{widetext}
\begin{equation}
\begin{aligned}
X^{T}_n(t)=&R_T[\cos\theta_{tq}\cos\phi_{tq}\sin(\alpha_{Tn}-\alpha_{T0}')+\sin\phi_{tq}\cos(\alpha_{Tn}-\alpha_{T0}')] \\
&+\frac 12 e_TR_T[\sin\phi_{tq}(\cos 2(\alpha_{Tn}-\alpha_{T0}')-3)+\cos\theta_{tq}\cos\phi_{tq}\sin 2(\alpha_{Tn}-\alpha_{T0}')]\\
&+\frac 14e^2_TR_T[-6\sin\phi_{tq}\cos(\alpha_{Tn}-\alpha_{T0}') \sin^2(\alpha_{Tn}-\alpha_{T0}')+\cos\theta_{tq}\cos\phi_{tq}(3\cos 2(\alpha_{Tn}-\alpha_{T0}')-1)]\\
&+R\cos(\alpha_T-\alpha_{T0})+\frac 12 eR(\cos 2(\alpha_T-\alpha_{T0})-3)-\frac 32 e^2R\cos(\alpha_T-\alpha_{T0})\sin^2(\alpha_T-\alpha_{T0})+O(e^3,e^3_1),
\end{aligned}
\end{equation}
% \end{widetext}

% \begin{widetext}
\begin{equation}
\begin{split}
Y^{T}_n(t)=&R_T[\cos\theta_{tq}\sin\phi_{tq}\sin(\alpha_{Tn}-\alpha_{T0}')-\cos\phi_{tq}\cos(\alpha_{Tn}-\alpha_{T0}')] \\
&+\frac 12 e_TR_T[-\cos\phi_{tq}(\cos 2(\alpha_{Tn}-\alpha_{T0}')-3)+\cos\theta_{tq}\sin\phi_{tq}\sin 2(\alpha_{Tn}-\alpha_{T0}')]\\
&+\frac 14e^2_TR_T[6\cos\phi_{tq}\cos(\alpha_{Tn}-\alpha_{T0}')\sin^2(\alpha_{Tn}-\alpha_{T0}')+\\
&\cos\theta_{tq}\sin\phi_{tq}\sin(\alpha_{Tn}-\alpha_{T0}')(3\cos 2(\alpha_{Tn}-\alpha_{T0}')-1)] \\
&+R\sin(\alpha_T-\alpha_{T0})+\frac 12 eR\sin 2(\alpha_T-\alpha_{T0})+\frac 14 e^2R\sin(\alpha_T-\alpha_{T0})(e\cos 2(\alpha_T-\alpha_{T0})-1)+O(e^3,e^3_1),
\end{split}
\end{equation}
% \end{widetext}

% \begin{widetext}
\begin{equation}
\begin{split}
Z^{T}_n(t)=&-R_T\sin\theta_{tq}\sin(\alpha_{Tn}-\alpha_{T0}')-\frac 12 e_TR_T\sin\theta_{tq}\sin 2(\alpha_{Tn}-\alpha_{T0}')\\
&-\frac 14e^2_TR_T\sin\theta_{tq}\sin(\alpha_{Tn}-\alpha_{T0}') \times[3\cos 2(\alpha_{Tn}-\alpha_{T0}')-1],
\end{split}
\end{equation}
\end{widetext}
where ($\theta_{tq}=-4.7^\circ$, $\phi_{tq}=120.5^\circ$) is the location of the source RX J0806.3+1527,
$e_T$ is the eccentricity of the orbit which we set to be 0 in this paper,
$e=0.0167$ is the eccentricity of the Earth's orbit, $R=1$ AU,
$\alpha_{T0}$ and $\alpha'_{T0}$ are initial phases, $\alpha_T=2\pi f_mt+\kappa_0$, $f_m=1/(365$ days),
$\alpha_{Tn}=2\pi f_{sc}t+2\pi(n-1)/3$.
For the detector C, the semimajor axis of the spacecraft's orbit
$R_T=3.7\times 10^6$ km, $f_{sc}=1/(365$ days).
For the detector C1, $R_T=1.7\times 10^5$ km, $f_{sc}=1/(365$ days).
For the detector C2, $R_T=1.7\times 10^5$ km, $f_{sc}=1/(3.65$ days).

\subsection{The orbits for the detectors R and R1}

The orbit $\vec{r}^L_n(t)=(X^{L}_n(t),Y^{L}_n(t),Z^{L}_n(t))$ in the ecliptic coordinate system is
% \begin{widetext}
\begin{gather}
X^{L}_n(t)=R(\cos\alpha_{Ln}+e_L)\cos\epsilon\cos\theta_{Ln}\nonumber\\
-R\sqrt{1-e_L^2}\sin\alpha_{Ln}\sin\theta_{Ln},\\
Y^{L}_n(t)=R(\cos\alpha_{Ln}+e_L)\cos\epsilon\sin\theta_{Ln}\nonumber\\
+R\sqrt{1-e_L^2}\sin\alpha_{Ln}\cos\theta_{Ln},\\
Z^{L}_n(t)=R(\cos\alpha_{Ln}+e_L)\sin\epsilon,
\end{gather}
% \end{widetext}
where $e_L=(1+\frac{2}{\sqrt{3}}\sigma+\frac{4}{3}\sigma^2)^\frac{1}{2}-1$,
$\epsilon=\arctan[\sigma/(1+\sigma/\sqrt{3})]$,
$\sigma=L/(2R)$, $R\approx 1$ AU,
$\theta_{Ln}=2\pi(n-1)/3$, $\alpha_{Ln}+e_{L}\sin\alpha_{Ln}=2\pi f_m t-2\pi(n-1)/3-\alpha_{L0}$
and $\alpha_{L0}$ is the initial phase. 
The arm length $L=3.7\times10^9$ m for the detector R 
and $L=1.7\times10^8$ m for the detector R1.

\section{THE EFFECT DIFFERENT CONSTELLATION FOR SOURCES WITH FIXED AMPLITUDE}
\label{results}
\begin{figure}[htp]
	\centering
	\includegraphics[width=0.9\columnwidth]{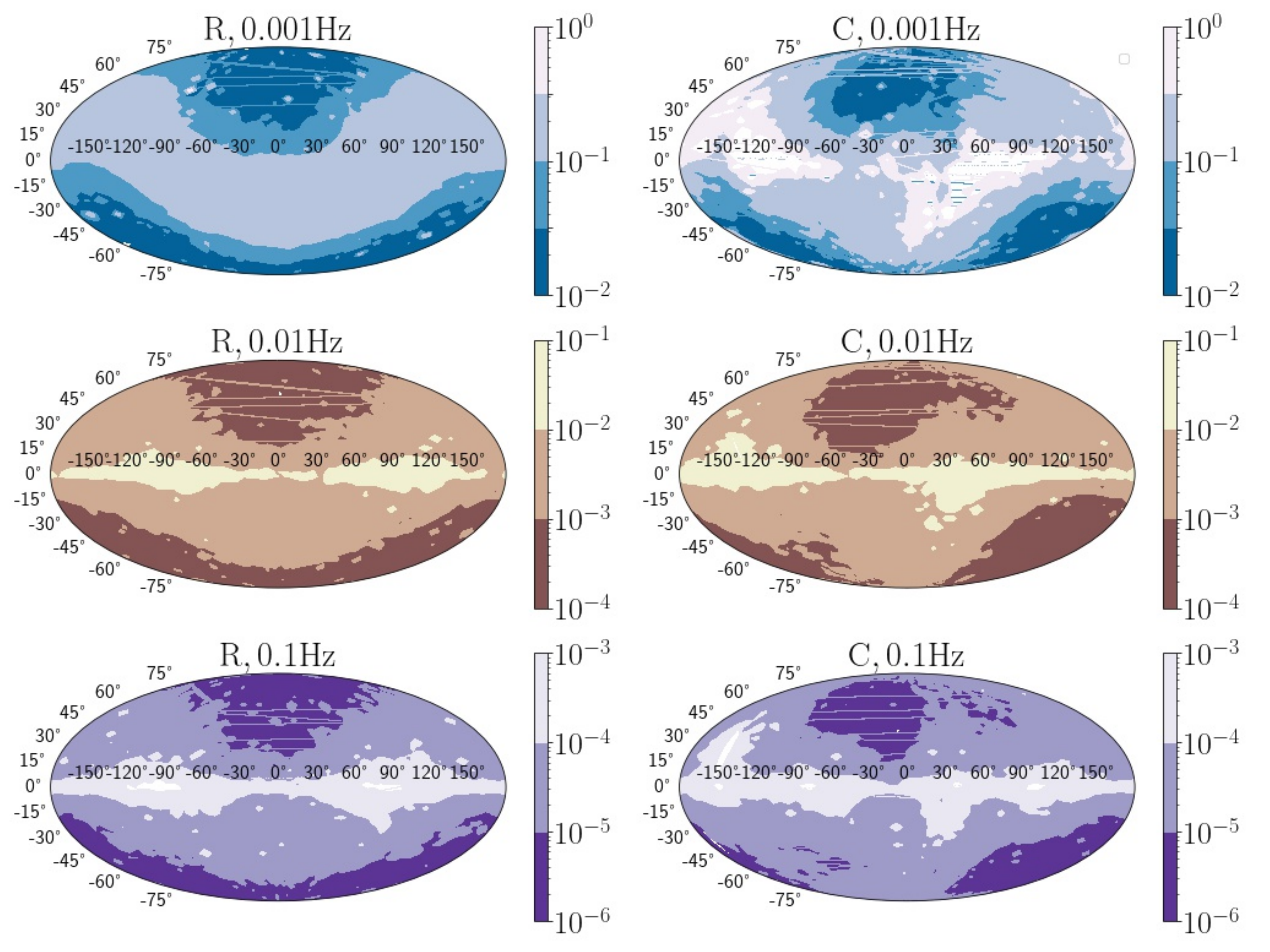}
	\caption{The sky map of angular resolutions $\Delta \Omega_S$  for
the fiducial detectors R and C. 
For the same frequency, the sources have the same amplitude.
The left panel is for the detector R and the right panel is for the detector C. From top to bottom, the frequencies of
monochromatic sources are $10^{-3}$ Hz, $10^{-2}$ Hz and $10^{-1}$ Hz.}
\label{rcfig2}
\end{figure}

\begin{figure*}[htp]
\centering
\includegraphics[width=0.95\textwidth]{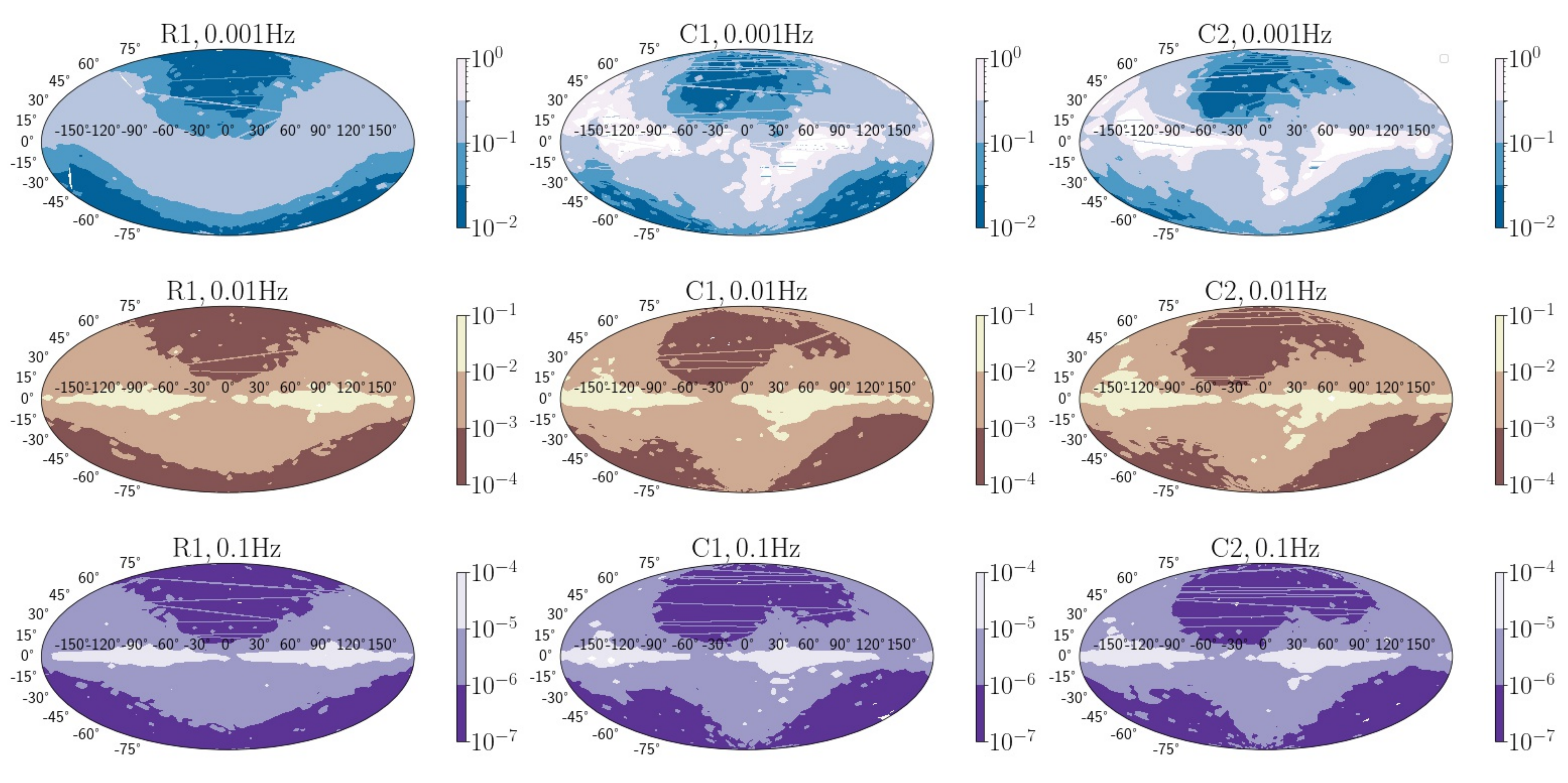}
\caption{The sky map of angular resolutions $\Delta \Omega_S$ for the detectors R1, C1 and C2. 
	For the same frequency, the sources have the same amplitude
	The left panel is for the detector R1,
the middle panel is for the detector C1 and the right panel is for the detector C2. From top to bottom, the frequencies are $10^{-3}$ Hz, $10^{-2}$ Hz and $10^{-1}$ Hz.}
\label{R1C1fig2}
\end{figure*}

In this section, instead of fixing the SNR, 
we fix the amplitude of the sources, i.e.,
we consider sources with the same distance and masses for different detectors.
To consider monochromatic GWs with the frequency ($10^{-1}, 10^{-2}, 10^{-3}$) Hz,
we take the sources to be equal mass binary systems with the total mass ($6 ,3\times10^2,10^4$) $M_{\odot}$ and the distance ($2.3, 1.3\times10^3, 10^4$) Mpc, respectively.
The noise power $S_n(f_0)$ is chosen as ($3.6\times10^{-41}, 3.6\times10^{-41}, 4.1\times10^{-41}$) at the frequency ($10^{-1}, 10^{-2}, 10^{-3}$), respectively.
The sky map of angular resolutions for the detectors R, C, R1, C1 and C2 are shown in Figs. \ref{rcfig2} and \ref{R1C1fig2} which are consistent with our conclusions (SNR=7).

\end{document}